# Structuring research methods and data with the research object model: genomics workflows as a case study

Kristina M Hettne[1]
Email: k.m.hettne@lumc.nl

Harish Dharuri[1]
Email: h.k.dharuri@lumc.nl

Jun Zhao[3]
Email: j.zhao5@lancaster.ac.uk

Katherine Wolstencroft[2,6]
Email: k.j.wolstencroft@liacs.leidenuniv.nl

Khalid Belhajjame[2]
Email: khalidb@cs.man.ac.uk

Stian Soiland-Reyes[2]
Email: soiland-reyes@cs.manchester.ac.uk

Eleni Mina[1]
Email: E.Mina@lumc.nl

Mark Thompson[1]
Email: M.Thompson@lumc.nl

Don Cruickshank[3]
Email: dgc@ecs.soton.ac.uk

Lourdes Verdes-Montenegro[5]
Email: lourdes@iaa.es

Julian Garrido[5]
Email: jgarrido@iaa.es

David de Roure[3]
Email: david.deroure@oerc.ox.ac.uk

Oscar Corcho[4]
Email: ocorcho@fi.upm.es

Graham Klyne[3]
Email: graham.klyne@oerc.ox.ac.uk

Reinout van Schouwen[1]
Email: reinout@gmail.com

Peter A C 't Hoen[1]
Email: P.A.C._t_Hoen@lumc.nl

Sean Bechhofer[2]
Email: sean.bechhofer@cs.man.ac.uk

Carole Goble[2]
Email: carole.goble@manchester.ac.uk

Marco Roos[1*]
* Corresponding author
Email: m.roos@lumc.nl

[1] Department of Human Genetics, Leiden University Medical Center, Leiden, The Netherlands

[2] School of Computer Science, University of Manchester, Manchester, UK

[3] Department of Zoology, University of Oxford, Oxford, UK

[4] Ontology Engineering Group, Universidad Politécnica de Madrid, Madrid, Spain

[5] Instituto de Astrofísica de Andalucía, Granada, Spain

[6] Leiden Institute of Advanced Computer Science, Leiden University, Leiden, The Netherlands

# Abstract

## Background

One of the main challenges for biomedical research lies in the computer-assisted integrative study of large and increasingly complex combinations of data in order to understand molecular mechanisms. The preservation of the materials and methods of such computational experiments with clear annotations is essential for understanding an experiment, and this is increasingly recognized in the bioinformatics community. Our assumption is that offering means of digital, structured aggregation and annotation of the objects of an experiment will provide necessary meta-data for a scientist to understand and recreate the results of an experiment. To support this we explored a model for the semantic description of a workflow-centric Research Object (RO), where an RO is defined as a resource that aggregates other resources, e.g., datasets, software, spreadsheets, text, etc. We applied this model to a case study where we analysed human metabolite variation by workflows.

## Results

We present the application of the workflow-centric RO model for our bioinformatics case study. Three workflows were produced following recently defined Best Practices for workflow design. By modelling the experiment as an RO, we were able to automatically query the experiment and answer questions such as "which particular data was input to a particular workflow to test a particular hypothesis?", and "which particular conclusions were drawn from a particular workflow?".

**Conclusions**

Applying a workflow-centric RO model to aggregate and annotate the resources used in a bioinformatics experiment, allowed us to retrieve the conclusions of the experiment in the context of the driving hypothesis, the executed workflows and their input data. The RO model is an extendable reference model that can be used by other systems as well.

**Availability**

The Research Object is available at http://www.myexperiment.org/packs/428

The Wf4Ever Research Object Model is available at http://wf4ever.github.io/ro

# Background

One of the main challenges for biomedical research lies in the integrative study of large and increasingly complex combinations of data in order to understand molecular mechanisms, for instance to explain the onset and progression of human diseases. Computer-assisted methodology is needed to perform these studies, posing new challenges for upholding scientific quality standards for the reproducibility of science. The aim of this paper is to describe how the research data, methods and metadata related to a workflow-centric computational experiment can be aggregated and annotated using standard Semantic Web technologies, with the purpose of helping scientists performing such experiments in meeting requirements for understanding, sharing, reuse and repurposing.

The workflow paradigm is gaining ground in bioinformatics as the technology of choice for recording the steps of computational experiments [1-4]. It allows scientists to delineate the steps of a complex analysis and expose this to peers using workflow design and execution tools such as Taverna [5], and Galaxy [6], and workflow sharing platforms such as myExperiment [7] and crowdLabs [8]. In a typical workflow, data outputs are generated from data inputs via a set of (potentially distributed) computational tasks that are coordinated following a workflow definition. However, workflows do not provide a complete solution for aggregating all data and all meta-data that are necessary for understanding the full context of an experiment. Consequently, scientists often find it difficult (or impossible) to reuse or repurpose existing workflows for their own analyses [9]. In fact, insufficient meta-data has been listed as one of the main causes of workflow decay in a recent study of Taverna workflows on myExperiment [9]. Workflow decay is the term used when the ability to re-execute a workflow after its inception has been compromised.

We will be able to better understand scientific workflows if we are able to capture more relevant data and meta-data about them; including the purpose and context of the experiment, sample input and output datasets, and the provenance of workflow executions. Moreover, if we wish to publish and exchange these resources as a unit, we need a mechanism for aggregation and annotation that would work in a broad scientific community. Semantic Web technology seems a logic choice of technology, given its focus on capturing the meaning of data in a machine readable format that is extendable and supports interoperability. It allows defining a Web-accessible reference model for the annotation of the aggregation and the aggregated resources that is independent of how data are stored in repositories. Examples of other efforts where Semantic Web technology has been used for the biomedical data integration includes the Semantic Enrichment of the Scientific Literature (SESL) [10] and Open PHACTS [11] projects. We applied the recently developed Research Object (RO) family of tools and ontologies [12,13] to preserve the scientific assets and their annotation related to a computational experiment. The concept of the RO was first proposed as an abstraction for sharing research investigation results [14]. Later, the potential role for ROs in facilitating not only the sharing but also the reuse of results, in order to increase the reproducibility of these results, was envisioned [15]. Narrowing down to workflow-centric ROs, preservation aspects were explored in [16], and their properties as first class citizen structures that aggregate resources in a principled manner in [13]. We also showed the principle of describing a (text mining) workflow experiment and its results by Web Ontology Language (OWL) ontologies [17]. The OWL ontologies were custom-built, which we argue is now an unnecessary bottleneck for exchange and interoperability. These studies all contributed to the understanding and implementation of the concept of an RO, but the data used were preliminary, and the studies were focused on describing workflows with related datasets and provenance information, rather than from the viewpoint of describing a scientific experiment of which workflows are a component.

A workflow-centric RO is defined as a resource that aggregates other resources, such as workflow(s), provenance, other objects and annotations. Consequently, an RO represents the method of analysis and all its associated materials and meta-data [13,15], distinguishing it from other work mainly focusing on provenance of research data [18,19]. Existing Semantic Web frameworks are used, such as (i) the Object Exchange and Reuse (ORE) model [20]; (ii) the Annotation Ontology (AO) [21]; and (iii) the W3C-recommended provenance exchange models [22]. ORE defines the standards for the description and exchange of aggregations of Web resources and provides the basis for the RO ontologies. AO is a general model for annotating resources and is used to describe the RO and its constituent resources as well as the relationships between them. The W3C provenance exchange models enable the interchange of provenance information on the Web, and the Provenance Ontology (PROV-O) forms the basis for recording the provenance of scientific workflow executions and their results.

In addition, we used the minimal information model "Minim", also in Semantic Web format, to specify which elements in an RO we consider "must haves", "should haves" and "could haves" according to user-defined requirements [23]. A checklist service subsequently queries the Minim annotations as an aid to make sufficiently complete ROs [24]. The idea of using a checklist to perform quality assessment is inspired by related checklist-based approaches in bioinformatics, such as the Minimum Information for Biological and Biomedical Information (MIBBI)-style models [25].

## Case study: genome wide association studies

As real-world example we aggregate and describe the research data, methods and metadata of a computational experiment in the context of studies of genetic variation in human metabolism. Given the potential of genetic variation data in extending our understanding of genetic diseases, drug development and treatment, it is crucial that the steps leading to new biological insights can be properly recorded and understood. Moreover, bioinformatics approaches typically involve aggregation of disparate online resources into complex data parsing pipelines. This makes this a fitting test case for an instantiated RO. The biological goal of the experiment is to aid in the interpretation of the results of a Genome-Wide Association Study (GWAS) by relating metabolic traits to the Single Nucleotide Polymorphisms (SNPs) that were identified by the GWAS. GWA studies have successfully identified genomic regions that dispose individuals to diseases (see for example [26], for a review see [27]). However, the underlying biological mechanisms often remain elusive, which led the research community to evince interest in genetic association studies of metabolites levels in blood (see for example [28-30]). The motivation is that the biochemical characteristics of the metabolite and the functional nature of affected genes can be combined to unravel biological mechanisms and gain functional insight into the aetiology of a disease. Our specific experiment involves mining curated pathway databases and a specific text mining method called concept profile matching [31,32].

In this paper we describe the current state of RO ontologies and tools for the aggregation and annotation of a computational experiment that we developed to elucidate the genetic basis for human metabolic variation.

# Methods

We performed our experiment using workflows developed in the open source Taverna Workflow Management System version 2.4 [5]. To improve the understanding of the experiment, we have added the following additional resources to the RO, using the RO-enabled myExperiment [33]: 1) the hypothesis or research question (what the experiment was designed to test); 2) a workflow-like sketch of the overall experiment (the overall data flow and workflow aims); 3) one or more workflows encapsulating the computational method; 4) input data (a record of the data that were used to reach the conclusions of an experiment); 5) provenance of workflow runs (the data lineage paths built from the workflow outputs to the originating inputs); 6) the results (a compilation of output data from workflow runs); 7) the conclusions (interpretation of the results from the workflows against the original hypothesis). Such an RO was then stored in the RO Digital Library [34]. RO completeness evaluation is checked from myExperiment with a tool implementing the Minim model [24]. Detailed description of the method follows.

## Workflow development

We developed three workflows for interpreting SNP-metabolite associations from a previously published genome-wide association study, using pathways from the KEGG metabolic pathway database [35] and Gene Ontology (GO) [36] biological process associations from text mining of PubMed. To understand an association of a SNP with a metabolite, researchers would like to know the gene in the vicinity of the SNP that is affected by the polymorphism. Then, researchers examine the functional nature of the gene and evaluate if it makes sense given the biochemical characteristics of the metabolite with which it is associated. This typically involves interrogation of biochemical pathway databases and mining existing literature. We would like to evaluate the utility of background knowledge present in the databases and literature in facilitating a biological interpretation of the statistically significant SNP-metabolite pairs. We do this by first determining the genes closest to the SNPs, and then reporting the pathways that these genes participate in. We implemented two main workflows for our experiment. The first one mines the manually curated KEGG database of metabolic pathway and gene associations that are available via the KEGG REST Services [37]. The second workflow mines the text-mining based database of associations between GO biological processes and genes behind the Anni 2.1 tool [31] that are available via the concept profile mining Web services [38]. We also created a workflow to list all possible concept sets in the concept profile database, to encourage reuse of the concept profile-based workflow for matching against other concept sets than GO biological processes. The workflows were developed following the 10 Best Practices for workflow design [39]. The Best Practices were developed to encourage re-use and prevent workflow decay, and briefly consists of the following steps:

1) Make a sketch workflow to help design the overall data flow and workflow aims, and to identify the tools and data resources required at each stage. The sketch could be created using for example flowchart symbols, or empty beanshells in Taverna.

2) Use modules, i.e. implement all executable components as separate, runnable workflows to make it easier for other scientists to reuse parts of a workflow at a later date.

3) Think about the output. A workflow has the potential to produce masses of data that need to be visualized and managed properly. Also, workflows can be used to integrate and visualise data as well as for analysing it, so one should consider how the results will be presented easily to the user.

4) Provide input and output examples to show the format of input required for the workflow and the type of output that should be produced. This is crucial for the understanding, validation, and maintenance of the workflow.

5) Annotate, i.e. choose meaningful names for the workflow title, inputs, outputs, and for the processes that constitute the workflow as well as for the interconnections between the components, so that annotations are not only a collection of static tags but capture the dynamics of the workflow. Accurately describing what individual services do, what data they consume and produce, and the aims of the workflow are all essential for use and reuse.

6) Make it executable from outside the local environment by for example using remote Web services, or platform independent code/plugins. Workflows are more reusable if they can be executed from anywhere. If there is need to use local services, library or tools, then the workflow should be annotated in order to define its dependencies.

7) Choose services carefully. Some services are more reliable or more stable than others, and examining which are the most popular can assist with this process.

8) Reuse existing workflows by for example searching collaborative platforms such as

myExperiment for workflows using the same Web service. If a workflow has been tried, tested and published, then reusing it can save a significant amount of time and resource.

9) Test and validate by defining test cases and implementing validation mechanisms in order to understand the limitations of workflows, and to monitor changes to underlying services.

10) Advertise and maintain by publishing the workflow on for example myExperiment, and performing frequent testing of the workflow and monitoring of the services used. Others can only reuse it if it is accessible and if it is updated when required, due to changes in underlying services.

## The RO core model

The RO model [12,13] aims at capturing the elements that are relevant for interpreting and preserving the results of scientific investigations, including the hypothesis investigated by the scientists, the data artefacts used and generated, as well as the methods and experiments employed during the investigation. As well as these elements, to allow third parties to understand the content of the RO, the RO model caters for annotations that describe the elements encapsulated by the ROs, as well as the RO as a whole. Therefore, two main constructs are at the heart of the RO model, namely aggregation and annotation. The work reported on in this article uses version 0.1 of the RO model, which is documented online [12].

Following myExperiment packs [7], ROs use the ORE model [20] to represent aggregation. Using ORE, an RO is defined as a resource that aggregates other resources, e.g., datasets, software, spreadsheets, text, etc. Specifically, the RO extends ORE to define three new concepts: i) ro:ResearchObject is a sub-class of ore:Aggregation which represents an aggregation of resources. ii) ro:Resource is a sub-class of ore:AggregatedResource representing a resource that is aggregated within an RO. iii) ro:Manifest is a sub-class of ore:ResourceMap, representing a resource that is used to describe the RO.

To support the annotation of ROs, their constituent resources, as well as their relationship, we use the Annotation Ontology [21]. Several types of annotations are supported by the Annotation Ontology, e.g., comments, textual annotations (classic tags) and semantic annotations, which relate elements of the ROs to concepts from underlying domain ontologies. We make use of the following Annotation Ontology terms: i) ao:Annotation, which acts as a handle for the annotation. ii) ao:annotatesResource, which represents the resource(s)/RO(s) subjects to annotation. iii) ao:body, which describes the target of the annotation. The body of the annotation takes the form of a set of Resource Description Framework (RDF) statements. Note that it is planned for later revisions of the RO model to use the successor of AO, the W3C Community Open Annotation Data Model (OA) [40]. For our purposes, OA annotations follows a very similar structure using oa:Annotation, oa:hasTarget and oa:hasBody.

## Support for workflow-centric ROs

A special kind of ROs that are supported by the model is what we call workflow-centric ROs, which, as indicated by the name, refer to those ROs that contain resources that are workflow specifications. The structure of the workflow in ROs is detailed using the wfdesc vocabulary

[41], and is defined as a graph in which the nodes refers to steps in the workflow, which we call wfdesc:Process, and the edges representing data flow dependencies, wfdesc:DataLink, which is a link between the output and input parameters (wfdesc:Parameter) of the processes that compose the workflow. As well as the description of the workflow, workflow centric ROs support the specification of the workflow runs, wfprov:WorkflowRun, that are obtained as a result of enacting workflows. A workflow run is specified using the wfprov ontology [42], which captures information about the input used to feed the workflow execution, the output results of the workflow run, as well as the constituent process runs, wfprov:ProcessRun, of the workflow run, which are obtained by invoking the workflow processes, and the input and outputs of those process runs.

## Support for domain-specific information

A key aspect of the RO model design is the freedom to use any vocabulary. This allows for inclusion of very domain-specific information about the RO if that serves the desired purpose of the user. We defined new terms under the name space roterms [43]. These new terms serve two main purposes. They are used to specify annotations that are, to our knowledge, not catered for by existing ontologies, e.g., the classes roterms:Hypothesis and roterms:Conclusion to annotate the hypothesis and conclusions part of an RO, and the property roterms:exampleValue to annotate an example value for a given input or output parameter given as an roterms:WorkflowValue instance. The roterms are also used to specify shortcuts that make the ontology easy to use and more accessible. For example, roterms:inputSelected associates a wfdesc:WorkflowDefinition to an ro:Resource to state that a file is meant to be used with a given workflow definition, without specifying at which input port or in which workflow run.

## Minim model for checklist evaluation

When building an RO in myExperiment users are provided with a mechanism of quality insurance by our so-called checklist evaluation tool, which is built upon the Minim checklist ontology [23,44] and defined using Web Ontology Language. Its basic function is to assess that all required information and descriptions about the aggregated resources are present and complete. Additionally, according to explicit requirements defined in a checklist, the tool can also assess the accessibility of those resources aggregated in an RO, in order to increase the trust on the understanding of the RO. The Minim model has four key components, as illustrated by Figure 1: 1) a Constraint, which associates a model (checklist) to use with an RO, for a specific assessment purpose, e.g. reviewing an RO containing sufficient information before being shared; 2) a Model, which enumerates of the set of requirements to be considered, which may be declared at levels of MUST, SHOULD or MAY be satisfied for the model as a whole; 3) a Requirement, which is the key part for expressing the concrete quality requirements to an RO , for example, the presence of certain information about an experiment, or liveness (accessibility) of a data server; 4) a Rule, which can be a SoftwareRequirementRule, to specify the software to be present in the operating environment, a ContentMatchRequirementRule, to specify the presence of certain pattern in the assessed data, or a DataRequirementRule, for specifying data resource to be aggregated in an RO.

**Figure 1 An overview of the Minim model**. An overview of the four components: a constraint, a model, a requirement, and a rule.

## RO digital library

While myExperiment acts mainly as front-end to users, the RO Digital Library [34] acts as a back-end, with two complementary storage components: a digital repository to keep the content, as a triple store to manage the meta-data content. The ROs in the repository can be accessed via a Restful API [45] or via a public SPARQL endpoint [46]. All the ROs created in the myExperiment.org are also submitted to the RO Digital Library.

## Workflow-centric RO creation process

Below we describe the steps that we conducted when creating the RO for our case study in an "RO-enabled" version of myExperiment [33]. The populated RO is intended to contain all the information required to re-run the experiment, or understand the results presented, or both.

## Creating an RO

The action of creating an RO consists of generating the container for the items that will be aggregated, and getting a resolvable identifier for it. In myExperiment the action of creating an RO is similar to creating a pack. We filled in a title and description of the RO at the point of creation and got a confirmation that the RO had been created and had been assigned a resolvable identifier in the RO Digital Library (Figure 2).

**Figure 2 Screenshots from myExperiment illustrating the process of creating a Research Object placeholder.** Before pressing the "create" button the user can enter a title and description (**A**), while pressing the "create" button will result in a placeholder Research Object with an identifier (**B**).

## Adding the experiment sketch

Using a popular office presentation tool, we made an experiment sketch and saved it as a PNG image. We then uploaded the image to the pack, selecting the type "Sketch". As a result, the image gets stored in the Digital Library and aggregated in the RO. In addition, an annotation was added to the RO to specify that the image is of type "Sketch". A miniature version of the sketch is shown within the myExperiment pack (Figure 3).

**Figure 3 Workflow sketch.** A workflow sketch showing that our experiment follows two paths to interpret genome wide association study results: matching with concept profiles and matching with KEGG pathways.

## Adding the hypothesis

To specify the hypothesis, we created a text file that describes the hypothesis, and then upload it to the pack as type "Hypothesis". The file gets stored in the Digital Library and aggregated in the RO, this time annotated to be of type "Hypothesis".

## Adding workflows

We saved the workflow definitions to files and uploaded them to the pack as type "Workflow". MyExperiment then automatically performed a workflow-to-RDF transformation in order to extract the workflow structure according to the RO model, which includes user descriptions and metadata created within the Taverna workbench. The descriptions and the extracted structure gets stored in the RO Digital Library and associated with the workflow files as annotations.

## Adding the workflow input file

The data values were stored in files that were then uploaded into the pack as "Example inputs". Such files gets stored in the RO Digital Library and aggregated in the RO, and as "Example inputs".

## Adding the workflow provenance

Using the Taverna-Prov [47] extension to Taverna, we exported the workflow run provenance to a file that we uploaded to the pack as type "Workflow run". Similar to other resources, the provenance file gets stored in the digital library with the type "Workflow run", however as the file is in the form of RDF according to the wfprov [42] and W3C PROV-O [22] ontologies, it is also integrated into the RDF store of the digital library and available for later querying.

## Adding the results

We made a compilation of the different workflow outputs to a result file in table format, uploaded to the pack as type "Results". The file gets stored in the digital library and aggregated in the RO, annotated to be of the type "Results".

## Adding the conclusions

To specify the hypothesis, we created a text file that describes the hypothesis, and then uploaded it to the pack as type "Hypothesis". The file gets stored in the digital library and aggregated in the RO, annotated to be of type "Conclusions".

## Intermediate step: checklist evaluation

At this point we checked how far we were from satisfying the Minim model, and were informed by the tool that the RO now fully satisfies the checklist (Figure 4).

**Figure 4 Screenshot of the results from the second check with the checklist evaluation service.** The results from checklist evaluation service show that the Research Object satisfies the defined checklist for a Research Object.

## Annotating and linking the resources

We linked the example input file to the workflows that used the file by the property "Input_selected" (Figure 5). In this particular case, both workflows have the same inputs but they need to be configured in different ways. This is described in the workflow description field in Taverna.

**Figure 5 Screenshot of the relationships in the RO in myExperiment.** The relationships between example inputs and workflows in the Research Object have been defined in myExperiment.

# Results

The RO for our experiment is the container for the items that we wished to aggregate. In terms of RDF, we first instantiated an ro:ResearchObject in an RO-enabled version of myExperiment [33]. We thereby obtained a unique and resolvable Uniform Resource Identifier (URI) from the RO store that underlies this version of myExperiment. In our experimental setup this was http://sandbox.wf4ever-project.org/rodl/ROs/Pack405/. It is accessible from myExperiment [48]. Each of the subsequent items in the RO was aggregated as an ro:Resource, indicating that the item is considered a constituent member of the RO from the point of view of the scientist (the creator of the RO).

## Aggregated resources

We aggregated the following items: 1) the hypothesis (roterms:Hypothesis): we hypothesized that SNPs can be functionally annotated using metabolic pathway information complemented by text mining, and that this will lead to formulating new hypotheses regarding the role of genomic variation in biological processes; 2) the sketch (roterms:Sketch) shows that our experiment follows two paths to interpret SNP data: matching with concept profiles and matching with Kyoto Encyclopedia of Genes and Genomes (KEGG) pathways (Figure 3); 3) the workflows (wfdesc:Workflow): Figure 6 shows the workflow diagram for the KEGG workflow and Figure 7 shows the workflow diagram for the concept profile matching workflow. In Taverna, we aimed to provide sufficient annotation of the inputs, outputs and the functions of each part of the workflow to ensure a clear interpretation and to ensure that scientists know how to replay the workflows using the same input data, or re-run them with their own data. We provided textual descriptions in Taverna of each step of the workflow, in particular to indicate their purpose within the workflow (Figure 8); 4) the input data (roterms:exampleValue) that we aggregated in our RO was a list of example SNPs derived from the chosen GWAS [28]; 5) the workflow run provenance (roterms:WorkflowRunBundle): a ZIP archive that contains the intermediate values of the workflow run, together with its provenance trace expressed using wfprov:WorkflowRun and subsequent terms from the wfprov ontology. We thus stored process information from the input of the workflow execution to its output results, including the information for each constituent process run in the workflow run, modelled as wfprov:ProcessRun. The run data is: 3 zip files containing 2090 intermediate values as separate files totalling 9.7 MiB, in addition to 5 MiB of provenance traces; 6) the results (roterms:Result) were compiled from the different workflow outputs to one results file (see result document in the RO [49] Additional file 1). For 15 SNPs it lists the associated gene name, the biological annotation from the GWAS publication, the associated KEGG pathway, and the most strongly associated biological process according to concept profile matching. Our workflows were able to compute a biological annotation from KEGG for 10 out of 15 SNPs and 15 from mining PubMed. All KEGG annotations and most text mining annotations corresponded to the annotations by Illig et al [28]. An important result of the text mining workflow was the SNP-annotation "rs7156144- stimulation of tumor necrosis factor production", which represents a hypothetical relation that to our knowledge was not reported before; 7) the conclusions (roterms:Conclusion): we concluded that our KEGG and text mining workflows were successful in retrieving biological annotations for significant SNPs from a GWAS experiment, and predicting novel annotations.

**Figure 6 Taverna workflow diagram for the KEGG workflow.** Blue boxes are workflow inputs, brown boxes are scripts, grey boxes are constant values, green boxes are Web services, purple boxes are Taverna internal services, and pink boxes are nested workflows.

**Figure 7 Taverna workflow diagram for the concept profile mining workflow.** Blue boxes are workflow inputs, purple boxes are Taverna internal services, and pink boxes are nested workflows.

**Figure 8 Taverna workflow annotation example.** An example of an annotation of the purpose of a nested workflow in Taverna.

As an example of our instantiated RO, Figure 9 provides a simplified view of the RDF graph that aggregates and annotates the KEGG mining workflow. It shows the result of uploading our Taverna workflow to myExperiment, as it initiated an automatic transformation from a Taverna 2 t2flow file to a Taverna 3 workflow bundle, while extracting the workflow structure and user descriptions in terms of the wfdesc model [41]. The resulting RDF document was aggregated in the RO and used as the annotation body of a ao:Annotation on the workflow, thus creating a link between the aggregated workflow file and its description in RDF. The Annotation Ontology uses named graphs for semantic annotation bodies. In the downloadable ZIP archive of an RO each named graph is available as a separate RDF document, which can be useful in current RDF triple stores that do not yet fully support named graphs. The other workflows were aggregated and annotated in the same way. The RO model further uses common Dublin Core vocabulary terms [50] for basic metadata such as creator, title, and description.

**Figure 9 Simplified diagram showing part of the Research Object for our experiment.** The Research Object contains the items that were aggregated by the "Research Object-enabled" version of myExperiment. Shown is the part of the RDF graph that aggregates and annotates the KEGG pathway mining workflow.

In some cases we manually inserted specified relations between the RO resources via the myExperiment user interface. An example is the link between input data and the appropriate workflow for cases when an RO has multiple workflows and multiple example inputs. In our case, both workflows have the same inputs, but they need to be configured in different ways. This was described in the workflow description field in Taverna which becomes available from an annotation body in the workflow upload process.

## Checking for completeness of an RO: application of the Minim model

We also applied Semantic Web technology for checking the completeness of our RO. We implemented a checklist for the items that we consider essential or desirable for understanding a workflow-based experiment by annotating the corresponding parts of the RO model with the appropriate term from the Minim vocabulary (Table 1). Thus, some parts were annotated as "MUST have" with the property minim:hasMustRequirement (e.g. at least one workflow definition), and others as "SHOULD have" with the property minim:hasShouldRequirement (e.g. the overall sketch of the experiment). The complete checklist document can be found online in RDF format [51] and in a format based on the spreadsheet description of the workflow [52]. We subsequently used a checklist service that evaluates if an RO is complete by executing SPARQL queries on the Minim mappings. The overall result is a summary of the requirement levels associated with the individual items; e.g. a missing MUST requirement is a more serious omission than a missing SHOULD (or COULD) requirement. We justified the less strict requirements for some items to accommodate cases when an RO is used to publish a method as such. We found that treating the requirement levels as mutually exclusive (hence not sub properties) simplifies the implementation of checklist evaluation, and in particular the generation of results when a checklist item is not satisfied.

**Table 1 RO items checklist**

| Research Object item | Requirement | RO ontology term |
|---|---|---|
| Hypothesis or Research question | Should | roterms:Hypothesis / roterms:ResearchQuestion |
| Workflow sketch | Should | roterms:Sketch |
| One or more workflows | Must | wfdesc:Workflow |
| Web services of the workflow | Must | wfdesc:Process |
| Example input data | Must | roterms:exampleValue |
| Provenance of workflow runs | Must | wfprov:WorkflowRun |
| Example results | Must | roterms:Result |
| Conclusions | Must | roterms:Conclusion |

RO items for a workflow-based experiment annotated with the appropriate term from the Minim vocabulary.

# Discussion

In this paper we explored the application of the Semantic Web encoded RO model to provide a container data model for preserving sufficient information for researchers to understand a computational experiment. We found that the model indeed allowed us to aggregate the necessary material together with sufficient annotation (both for machines and humans). Moreover, mapping of selected RO model artefacts to the Minim vocabulary allowed us to check if the RO was complete according to our own predefined criteria. The checklist service can be configured to accommodate different criteria. Research groups may have different views on what is essential, but also libraries or publishers may define their own standards, enabling partial automation of the process of checking a submission against specific instructions to authors. Furthermore, the service can be run routinely to check for workflow decay, in particular decay related to references that go missing.

In using the RO model, we sought to meet requirements for sharing, reuse and repurposing, as well as interoperability and reproducibility. This fits with current trends to enhance reproducibility and transparency of science (e.g. see [53-55]). Reproducibility in computational science has been defined as a spectrum [55], where a computational experiment that is described only by a publication is not seen as reproducible, while adding code, data, and finally the linked data and execution data will move the experiment towards full replication. Adhering to this definition, our RO-enabled computational experiment comes close to fulfilling the ultimate golden standard of full replication, but falls short because it has not been analyzed using independently collected data. The benefit offered by the RO in terms of reproducibility is that it provides a context (RO) within which an evaluation of reproducibility can be performed. It does this by providing an enumerated and closed set of resources that are part of the experiment concerned, and by providing descriptive metadata (annotations) that may be specific to that context. This is not necessarily the complete solution to reproducible research, but at least an incremental step in that direction.

We have used RDF as the underlying data model for exchanging ROs. One of advantages is the ability to query the data, which becomes clear when we want to answer questions about the experiment, such as: 1) Which conclusions were drawn from a given workflow?; 2) Which workflow (run) supports a particular conclusion and which datasets did it use as inputs?; 3) Which different workflows used the same dataset X as input?; 4) Who can be credited for creating workflows that use GWAS data? The answers for the first two questions can readily be found using a simple SPARQL [56] query. Figure 10 shows the SPARQL query and the results as returned by the SPARQL endpoint of the RO Digital Library. Note that in our case we got two result rows, one for each of the workflows that were used to confirm the hypothesis. We emphasize that queries could also be constructed to answer more elaborate questions such as question 3 and 4. Without adding any complexity to the query or the infrastructure, it is possible to query over the entire repository of research objects. This effectively integrates all meta-data of any workflow-based experiment that was uploaded to the RO Digital Library via myExperiment. When more ROs have become available that use the same annotations as described in this paper, then we can start sharing queries that can act as templates. We did not explore further formalization in terms of rejecting or accepting hypotheses, since formulating such a hypothesis model properly would be very domain specific, such as current efforts in neuromedicine [57]. However, the RO model does not exclude the possibility to do so.

**Figure 10 Screenshot showing a SPARQL query and its results.** Query to obtain a reference to the data that was used as input to our workflows and the conclusions that we drew from evaluating the workflow results.

## Applying the RO model in genomic working environments

An important criterion for our evaluation of the RO model and tools is that it should support researchers in preparing their digital methods and results for publication. We have shown that the RO model can be applied in an existing framework for sharing computational workflows (myExperiment). We used Taverna to create our workflows, and the wf4ever toolkit [58], including dLibra [59] that was extended with a triple store as a back end to store the ROs. The RO features of the test version of myExperiment that we used are currently under development for migration to the production version of myExperiment [60]. Creating an RO in the test version of myExperiment is not any different to a user than the action of creating a pack, completely hiding the creation of RDF objects under the hood. The difference lies in the support of the RO model, which allows the user to add data associated with a computational experiment in a structured way (a sketch representing the experimental setup, the hypothesis document, result files, etc.), and metadata in the form of annotations. Every piece of data in an RO can be annotated, either in a structured or machine-generated way like the automatic annotation of a wfdesc description of a workflow as provided by the workflow-to-RDF transformation service, or manually by the user at the time of resource upload, such as the annotation of an experiment overview as "Sketch". Since RO descriptions are currently not a pre-requisite to publishing workflow results in journal, we hope that this support and streamlining of the annotation process will act as an incentive for scientists to start using the RO technology.

The representation of an RO in myExperiment as presented in this paper should be seen as a proof-of-concept. Crucial elements of a computational experiment are handled, but there is room for improvement. For example, the hypothesis and conclusions are at the moment only shown as downloadable text files and the content and provenance of a workflow run is not shown to the user. We found that more tooling is needed to make practical use of the provenance trace. It is detailed and focus is on data lineage, rather than the biological meaning of the recorded steps. Nevertheless, we regard this raw workflow data as highly valuable as the true record of what exactly was executed. It allows introspection of the data lineage, such as which service was invoked with exactly which data. By providing this proof-of-concept and the RO model as a reference model, we hope to stimulate developers of other genomic working environments such as Galaxy [6] and Genome Space [61] to start implementing the RO model as well, thus enabling scientists to share their investigation results as a complete knowledge package. Similarly, workflow systems use different workflow languages [62,63], and by presenting the workflow-to-RDF transformation service that handles the t2flow serialization format to transform a workflow to an RO, we hope to encourage systems that use other workflow languages to develop similar services to transform their workflows to ROs. This would allow for a higher-level understanding of workflow-based experiments regardless of the type of workflow system used.

It should be noted that although our ROs fully capture the individual data items of individual steps within workflow runs, this approach is not applicable to all scientific workflows. In fact, we have since further developed the provenance support for Taverna so that larger pieces of data are only recorded as URI references and not bundled within the ZIP file. The Taverna workflow system already supports working with such references; however many bioinformatics Web services still only support working directly with values. When dealing with references, the workflow run data only capture the URI and its metadata, and full access to the run data therefore would also depend on the continued availability (or mirroring) of those referenced resources, and their consistency would therefore later need to be verified against metadata such as byte size and Secure Hash Algorithm checksums.

## Generalization to other domains

We acknowledge that apart from enabling the structured aggregation and annotation of digital ROs technically, scientists appreciate guidelines and Best Practices for producing high quality ROs. In fact, the minimal requirements for a complete RO that we implemented via the Minim model, were inspired by the 10 Best Practices that we defined for creating workflows [39]. An RO may be evaluated using different checklists for different purposes. A checklist description is published as linked data, and may be included in the RO, though we anticipate more common use will be for it to be published separately in a community web site. In our work to date, we have used checklist definitions published via Github (e.g. [64]), and are looking to create a collection of example checklist definitions to seed creation of checklists for different domains or purposes [65]. We envision that instructions to authors of ROs may differ between research communities, and publishers who wish to adopt RO technology for digital submissions may develop their own 'Instructions to Authors' for ROs. This could be implemented by different mappings of the Minim model.

## Related work

The RO model was implemented as a Semantic Web model to provide a general, domain-agnostic reference that can be extended by domain specific ontologies. For instance, while the RO model offers terms pertaining to experimental science such as "hypothesis" and "conclusion", extensions to existing models that also cover this area and are already in use in the life science domain could be considered. It is beyond the scope of this article to exhaustively review related ontologies and associated tools, but we wish to mention six that in our view are prime candidates to augment the RO family of ontologies and tools. The first is the Ontology for Biomedical Investigations (OBI) that aims to represent all phases of experimental processes, such as study designs, protocols, instrumentation, biological material, collected data and analyses performed on that data [66]. OBI is used for the ontological representation of the results of the Investigation-Study-Assay (ISA) metadata tools [67] that is the next on our list of candidates. ISA, developed by the ISA commons community, facilitates curation, management, and reuse of omics datasets in a variety of life science domains [68]. It puts spreadsheets at the heart of its tooling, making it highly popular for study capture in the omics domain [69]. The third candidate is the ontology for scientific experiments EXPO [70]. EXPO is defined by OWL-DL axioms and is grounded in upper ontologies. Its coverage of experiment terms is good, but we are unsure about its uptake by the community. Perhaps unfortunate for a number of good ontologies, we consider this an important criterion for interoperability. Four and five on our list relate to the annotation of Web Services (or bioinformatics operations in general): the EMBRACE Data and Methods (EDAM) ontology encompasses over 2200 terms for annotating tool or workflow functions, types of data and identifiers, application domains and data formats [71]. It is developed and maintained by the European Bioinformatics Institute and has been adopted for annotation of for instance the European Molecular Biology Open Software Suite. The myGrid-BioMoby ontology served as a starting point for the development of EDAM. This will facilitate the adoption of EDAM by for instance BioCatalogue,org and service-oriented tools such as Taverna, which would further broaden its user base and thereby its use for interoperability. The Semantic Automated Discovery and Integration (SADI) framework [72] takes semantic annotation of Web Services one step further. A SADI Web Service describes itself in terms of OWL classes, and produces and consumes instances of OWL classes. This enables instant annotation in a machine readable format when a workflow is built from SADI services. In addition, via a SADI registry suggestions can be made about which services to connect to which. SADI has clear advantages as an annotation framework. However, not all bioinformatics services are available as SADI services, while the conversion is not trivial without training in Semantic Web modelling. Therefore, SADI and RO frameworks could be strongly complementary for workflows that use a heterogeneous mix of service types. This would be further facilitated when both are linked to common ontologies such as EDAM. Finally, we highlight the recent development of models for microattribution and nanopublication that aim to provide a means of getting credit for individual assertions and making these available in a machine readable format [73-75]. Taking nanopublications as an example, we could "nanopublish" specific results from our experiment, such as the text mining-based association that we found between the SNP "rs7156144" and the biological process "stimulation of tumor necrosis factor production". In addition to an assertion, a nanopublication consists of provenance meta-data (to ensure trust in the assertion) and publication information (providing attribution to authors and curators). Nanopublication and RO complement each other in two ways. On the one hand, nanopublications can be used to publish and expose valuable results from workflows and included in the RO aggregate. On the other hand, an RO could be referenced as part of the provenance of a nanopublication, serving as a record of the method that led to assertion of the nanopublication. Similar to the nanopublication and microattribution models, the Biotea and Elsevier Smart Content Initiative data models also aim to model scientific results, but are focused on encapsulating a collection of information that are related to the results reported in publications [76,77]. The relationship between an RO and these datasets is not much different from an RO with a nanopublication statement. An RO can be referenced by, e.g. the Biotea dataset, by its URI, which can provide detailed experimental information or provenance information about the results described by the Biotea dataset. In the meanwhile, an RO can also reference a Biotea dataset or an Elsevier linked dataset.

Summarizing, the RO model provides a general framework with terms for aggregating and annotating the components of digital research experiments, by which it can complement related frameworks that are already used in the life science domain such as EXPO, OBI, ISA, EDAM, SADI and nanopublication. We observe that models are partly complementary and partly overlapping in scope. Therefore, we stimulate collaboration towards the development of complementarity frameworks. For instance, we initiated an investigation of the combination of ISA, RO, and Nanopublication as a basis for general guidelines for publishing digital research artefacts (Manuscript in preparation).

**Uptake by the research community**

Beyond the RO presented in this paper, the RO model has been used to generate ROs within the domains of musicology [78] and astronomy using AstroTaverna [79]. In addition, we recently explored how an RO could be referenced as part of the provenance of nanopublications of genes that are differentially expressed in Huntington's Disease (HD) with certain genomic regions [80,81]. The results from the in silico analysis of the differentially expressed genes were obtained from a Taverna data integration workflow and the RO itself was stored in the Digital Library. Using the PROV-O ontology, the nanopublication provenance was modelled to link to the workflow description in the RO. Since the RO was mostly automatically generated by the procedure described in this paper, the nanopublication refers to detailed provenance information without requiring additional modelling effort. To encourage further uptake by the research community we have developed the Web resource ResearchObject.org [82]. ResearchObject.org lists example ROs [83], presents the ongoing activities of the open RO community, and gathers knowledge about related developments and adoptions.

# Conclusions

Applying the workflow-centric RO model and associated models such as Minim provides a digital method to increase the understanding of bioinformatics experiments. Crucial meta-data related to the experiment is preserved in a Digital Library by structured aggregation and annotation of hypothesis, input data, workflows, workflow runs, results, and conclusions. The Semantic Web representation provides a reference model for life scientists who perform computational analyses and for systems that support this, and can complement related annotation frameworks that are already in use in the life science domain.

# Competing interests

The authors declare that they have no competing interests.

# Authors' contributions

KMH participated in the design and the coordination of the study, participated in the design and creation of the workflows and the Web services used in the text-mining workflow, created the Research Object, and drafted the manuscript. HKD participated in the design of the study, participated in the design and creation of the workflows, and helped to draft the manuscript. JZ, KB, SSR, OC, GK, SB participated in the design of the semantic models and helped to draft the manuscript. KW, CG, LVM, JG, DR, PBH participated in the design of the study and helped to draft the manuscript. CG also prepared and co-supervised the work on Minim. EM performed the connection to the nanopublication model and helped to draft the manuscript. MT designed and performed the SPARQL queries and helped to draft the manuscript. DC implemented the requirements for creating a Research Object in myExperiment and helped to draft the manuscript. RS designed and implemented the web services used by the text-mining workflow. MR conceived of the study, participated in its design and coordination, and helped to draft the manuscript. All authors read and approved the final manuscript.

# Acknowledgements

The research reported in this paper is supported by the EU Wf4Ever STREP project (270129) funded under EU FP7 (ICT-2009.4.1), the EP/G026238/1 EPSRC project myGrid: A Platform for e-Biology Renewal, the IMI-JU project Open PHACTS (grant agreement n 115191), and grants received from the Netherlands Bioinformatics Centre (NBIC) under the BioAssist program.
We gratefully acknowledge Matt Gamble for his advice on the Minim model.

# Additional file

### Additional_file_1 as XLS
**Additional file 1** Research Object results. KEGG and Concept Profile Analysis comparison results.

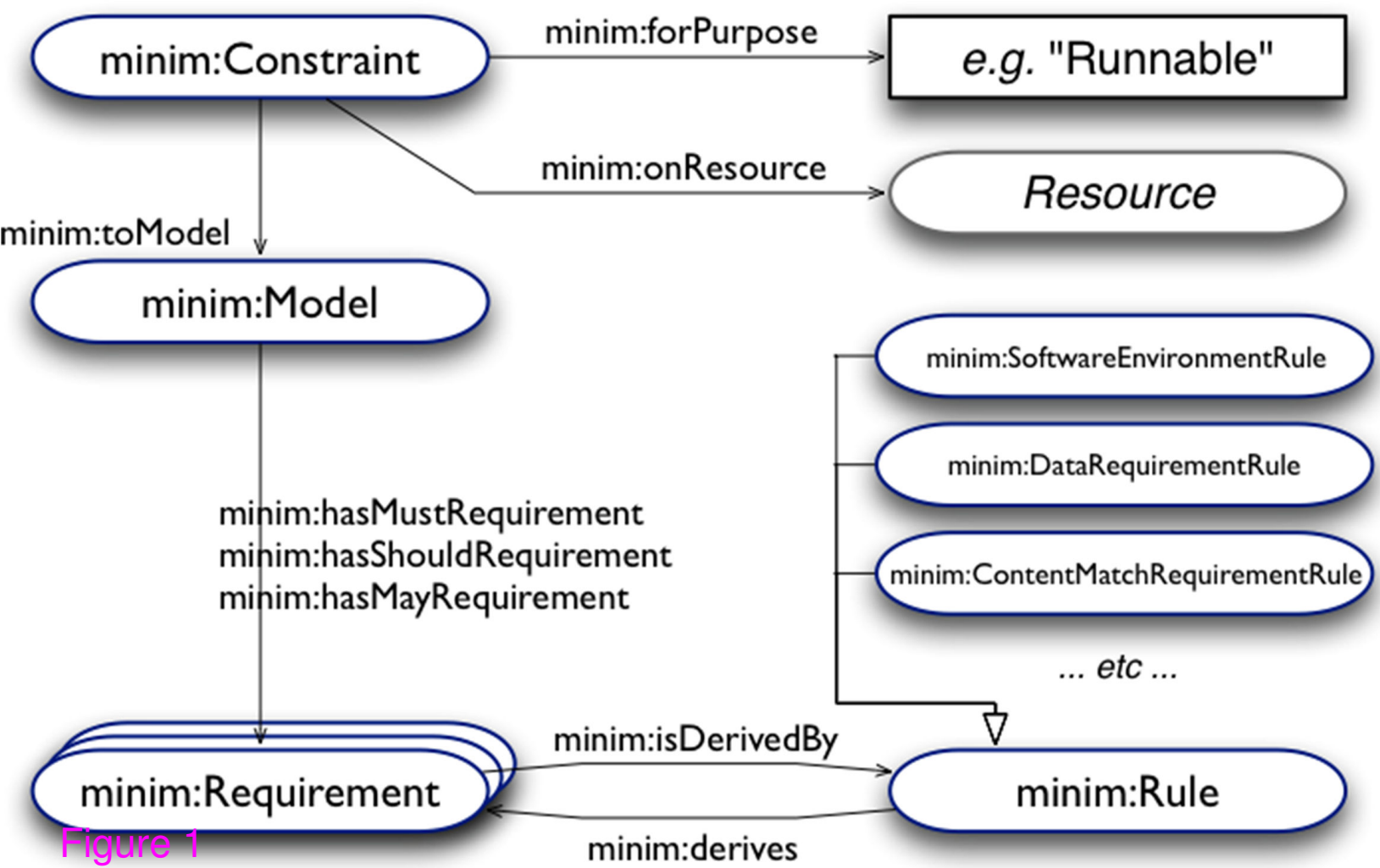

Figure 1

Figure 2

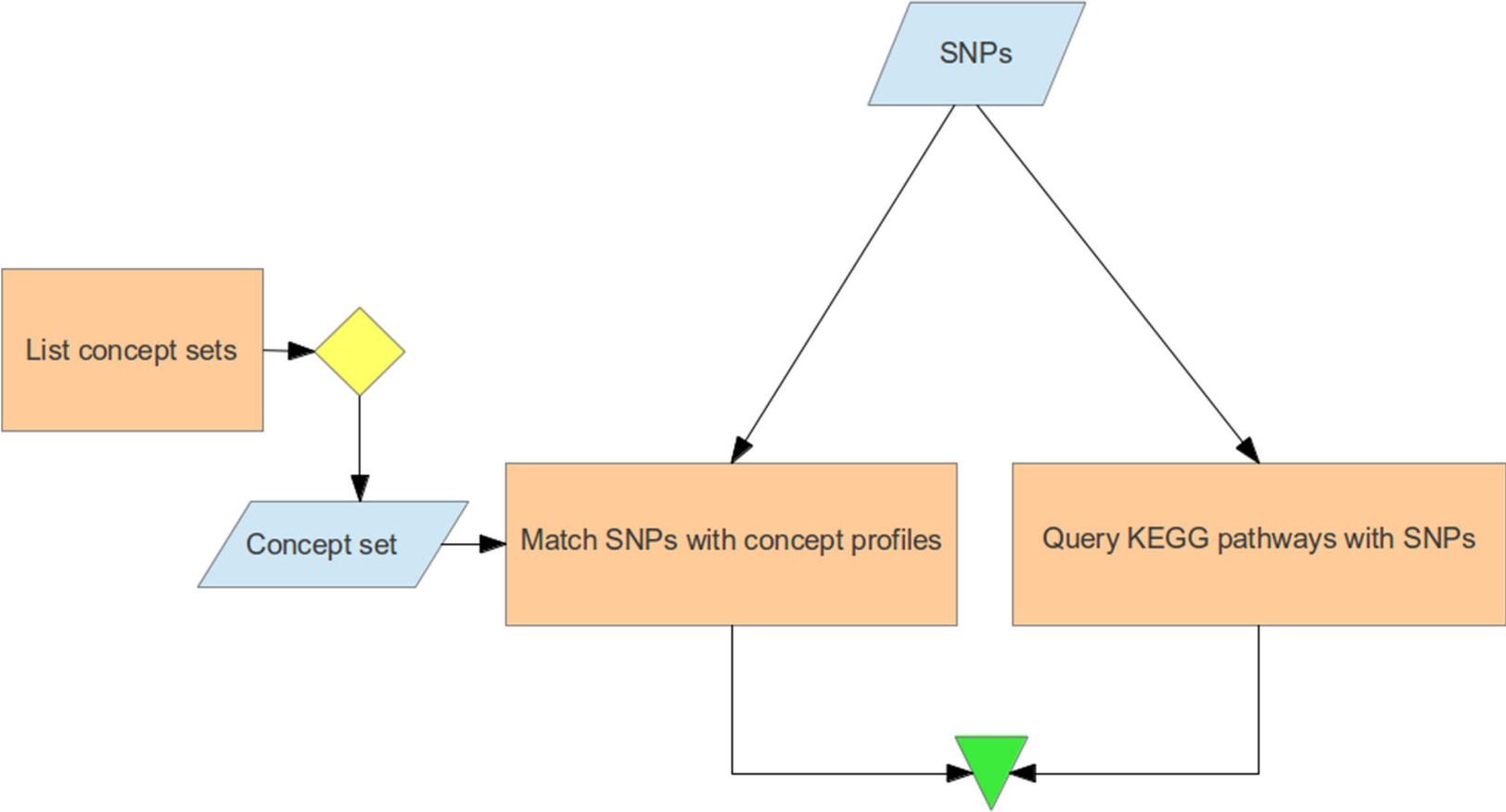

# Interpreting GWAS results with pathways and text mining

This pack contains workflows for interpreting SNP associations from a GWAS on human metabolite variation, using pathways from the KEGG metabolic pathway database and Gene Ontology biological process associations from text mining.

✓ **Target [Pack405](#) *fully satisfies* checklist for *ready-to-release*.**

- ✓ Experiment hypothesis is present
- ✓ Workflow design sketch is present
- ✓ All workflow definitions are accessible
- ✓ All web services used by workflows are accessible
- ✓ Input data is present
- ✓ Experiment conclusions are present

[Wf4Ever project](#)

Figure 4

Figure 5

Figure 6

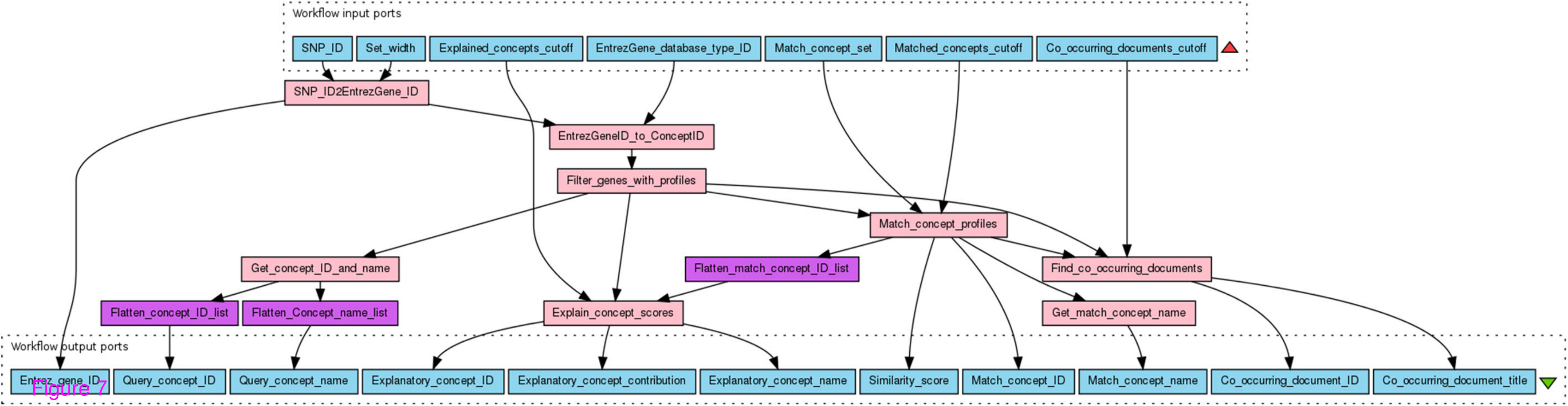

Figure 7

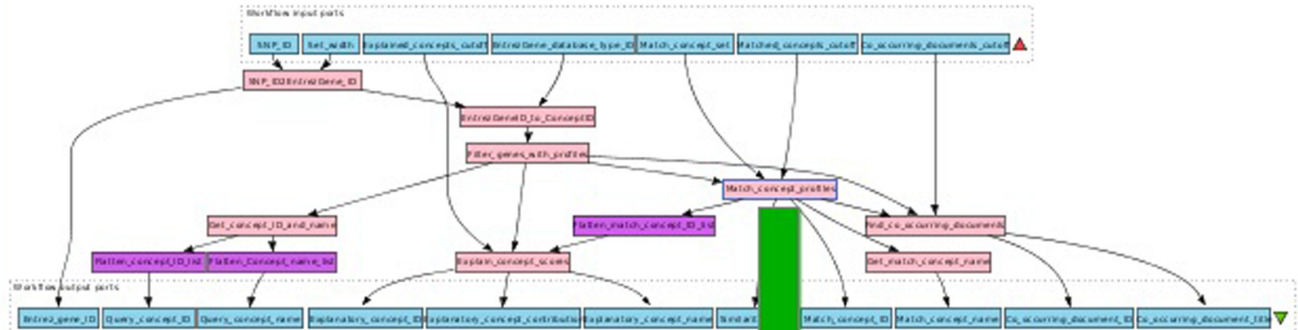

Figure 8

Figure 9

```sparql
PREFIX myRO: <http://sandbox.wf4ever-project.org/rodl/ROs/Pack405/>
PREFIX ro: <http://purl.org/wf4ever/ro#>
PREFIX ore: <http://www.openarchives.org/ore/terms/>
PREFIX roterms: <http://purl.org/wf4ever/roterms#>
PREFIX rdfs: <http://www.w3.org/2000/01/rdf-schema#>

SELECT * WHERE {
myRO: a ro:ResearchObject ;
  ore:aggregates ?hypothesis, ?workflow,
                 ?results, ?conclusions .
?hypothesis a roterms:Hypothesis .
?workflow roterms:inputSelected ?input .
?results a roterms:Results .
?conclusions a roterms:Conclusions .
}
```

| | hypothesis | input | workflow | results | conclusions |
|---|---|---|---|---|---|
| 1 | myRO:hypothesis.txt | myRO:top_snps_to_annotate_input.txt | myRO:GWAStoConcept.t2flow | myRO:kegg_cp_comparison_results.xls | myRO:conclusions.txt |
| 2 | myRO:hypothesis.txt | myRO:top_snps_to_annotate_input.txt | myRO:GWAStoPathway.t2flow | myRO:kegg_cp_comparison_results.xls | myRO:conclusions.txt |

Figure 10

**Additional files provided with this submission:**

Additional file 1: 7349194119911521_add1.xls, 9K
http://www.jbiomedsem.com/imedia/3679434621390084/supp1.xls